\documentclass[aps,twocolumn,prb,amsmath,amssymb,showpacs,superscriptaddress]{revtex4}
\usepackage{graphicx,multirow,array}
\usepackage{color}

\begin{document}
\title{The three-dimensional Heisenberg spin glass under a weak random
  anisotropy.}

\author{V. Martin-Mayor}
\affiliation{Departamento de F\'{\i}sica Te\'orica I, Facultad de Ciencias
  F\'{\i}sicas, Universidad Complutense, 28040 Madrid, Spain.}
\affiliation{Instituto de Biocomputaci\'on y F\'{\i}sica de Sistemas Complejos
  (BIFI), Corona de Arag\'on 42, Zaragoza 50009, Spain.}

\author{S. Perez-Gaviro} 
\affiliation{Dipartimento di Fisica, INFM and INFN, Universit\`a di
  Roma {\it La Sapienza}, Ple. A. Moro 2, 00185 Roma, Italy.}
\affiliation{Instituto de Biocomputaci\'on y F\'{\i}sica de Sistemas Complejos
  (BIFI), Corona de Arag\'on 42, Zaragoza 50009, Spain.}

\date{\today}

\begin{abstract}
We perform a finite size scaling study of the three-dimensional Heisenberg
spin glass in the presence of weak random anisotropic interactions, up to
sizes $L=32$. Anisotropies have a major impact on the phase transition. The
chiral-glass susceptibility does not diverge, due to a large anomalous
dimension. It follows that the anisotropic spin-glass belongs to a
Universality Class different from the isotropic model, which questions the
applicability of the chirality scenario.
\end{abstract}


\maketitle

\section{INTRODUCTION}

Spin glasses (SG's) are disordered magnetic alloys, widely regarded as
paradigmatic complex systems.\cite{SGINTRO} The degree of anisotropy
in the magnetic interactions determines whether a particular alloy is
classified as a Heisenberg or an Ising SG (Ising corresponds to a
limit of strong anisotropy). Experimentally, anisotropies affect
significantly the glassy response to external magnetic fields and the
behavior under cooling protocols.\cite{BERT04}

Theorists have privileged the study of the Ising limit, in spite of
the fact that canonical SG's, e.g. CuMn or AgMn, should be rather
regarded as Heisenberg, with weak anisotropic interactions. Indeed,
complications arise in the Heisenberg case. In addition to the
standard SG ordering, Heisenberg systems show as well a chiral-glass
(CG) phase, where {\em chiralities} order~\cite{VILLAIN} (chiralities,
also named vorticities, reflect the handedness of the non-collinear
spin ordering pattern, see definitions below).

Probably motivated by failures in early numerical attempts~\cite{Tc0}
to find a standard SG phase for Heisenberg systems, Kawamura proposed
a {\em chirality scenario}, expected to hold for most experimental
systems.\cite{CHIRALSCEN_KAWAMURA} In the ideal, fully isotropic
case, the standard SG critical temperature $T_\mathrm{SG}$ would be
strictly zero, while chiralities would order at $T_\mathrm{CG}>0$
(spin-chirality decoupling). Yet, anisotropic interactions (either
dipolar, pseudo-dipolar or
Dzyaloshinskii-Moriya\cite{LEVYFERT,BRAYMOORE}), albeit small, are
unavoidable in experimental samples. Hence, the scenario includes a
{\em decoupling-recoupling} hypothesis: weak random anisotropic
interactions would recouple spins and chiralities so that
$T_\mathrm{CG}=T_\mathrm{SG}>0$. Indeed, the numerical work available at
the time indicated that very small amounts of anisotropy lead to
$T_\mathrm{SG}>0$.\cite{MATSUBARA}

CG ordering may be experimentally investigated through the anomalous
Hall effect. Due to spin-orbit interaction and the spin polarization
of the conduction electrons, the anomalous Hall resistivity picks
contributions proportional to the CG order parameter and to its
corresponding non-linear susceptibility.\cite{AHE_TAKA,AHE_KAWA}  The
effectiveness of this tool to study non-coplanar orderings has been
demonstrated in manganites,\cite{AHE-MANG} and in a geometrically
frustrated pyrochlore ferromagnet.\cite{AHE-SC}

The effect of anisotropies on the critical behavior was considered by
Bray and Moore,\cite{BRAYMOORE} before the question of chiral
ordering was raised. They predicted that these systems belong to the
Ising SG's Universality Class, no matter the kind of anisotropic
interactions. However, in their analysis the assumption was made that
$T_\mathrm{SG}\!=\!0$ in the isotropic limit (this assumption seemed
plausible at the time, although we now know that it is incorrect).

Recent theoretical work has shown that the chirality scenario needs
some revision.  New simulation algorithms (allowing to thermalize at
lower temperatures than pioneering work~\cite{Tc0}), combined with
modern finite-size scaling (FSS)
methods,\cite{BALLESTEROS,QUOTIENTS,VICTORAMIT} have provided
conclusive evidence for a standard SG ordering with $T_\mathrm{SG}>0$
for purely isotropic
interactions.\cite{LEEYOUNG,LEEYOUNG2,HSGWE,HSG-LVSAP,VIET-KAWAMURA,OTROS,MATSHIENTA} Only some controversy remains on whether
$T_\mathrm{SG}$ is slightly smaller than
$T_\mathrm{CG}$,\cite{VIET-KAWAMURA} or rather the two are compatible
within errors.\cite{HSG-LVSAP} Interestingly enough, a modern-styled
study seems to be still lacking for the more realistic case of a
Heisenberg SG with small random anisotropy.

Here we show that small anisotropic interactions cause that, at
variance with the ideal case, the CG susceptibility no longer diverges
at $T_\mathrm{CG}$ (i.e. the anomalous dimension becomes
$\eta_\mathrm{CG}>2$).  In the Renormalization Group
framework,\cite{VICTORAMIT} anisotropy is a relevant
perturbation. Even if in an experimental sample anisotropies are
fairly small, the isotropic model is appropriate only for moderate
correlation length. Closer to the critical temperature, a new fixed
point rules (presumably in the Ising SG Universality Class, due to
spin-reversal symmetry).  A slow crossover~\cite{VICTORAMIT} from the
Heisenberg to the anisotropic fixed-point arises upon approaching the
phase transition.  We conjecture that this crossover
explains~\cite{FN1} experimental claims of
a non-trivial dependency of critical exponents on the anisotropy
strength.\cite{BERT04,CAMPBELLPETIT} Our results follow from a FSS
analysis of equilibrium Monte Carlo simulations on system-sizes up to
$L\!=\!32$. Data suggest that anisotropies cause a temperature range
in which chiralities order while spins do not
(i.e. $T_\mathrm{SG}<T_\mathrm{CG}$). However, due to the slow
crossover, further research will be needed to dismiss spin-chirality
recoupling.

The remaining part of this work is organized as follows. We define the model
and describe our numerical methods in Sect.~\ref{SECT:MODEL}. We address
thermal equilibration, a major issue in any spin-glass simulation, in
Sect.~\ref{SECT:EQUILIBRATION}. Our physical results are reported in
Sect.~\ref{SECT:RESULTS}. Finally, we give our results in
Sect.~\ref{SECT:CONCLUSIONS}.

\section{MODEL AND SIMULATIONS}\label{SECT:MODEL}

Since the main types of anisotropic interactions lead to the {\em
  same} effective replica Hamiltonian,\cite{BRAYMOORE} it is
numerically convenient to study short range (pseudo-dipolar)
interactions. Take the Edwards-Anderson model on a cubic lattice
of size $L$, with periodic boundary conditions. Heisenberg spins
occupy the lattice nodes $\boldsymbol{x}$ [$\vec
  S_{\boldsymbol{x}}=(S_{\boldsymbol{x}}^1, S_{\boldsymbol{x}}^2,
  S_{\boldsymbol{x}}^3)$, $\vec S_{\boldsymbol{x}}\cdot \vec
  S_{\boldsymbol{x}}=1\,$]. The Hamiltonian is~\cite{MATSUBARA}
\begin{equation}
  H=- \sum_{\langle \boldsymbol{x},\boldsymbol{y}\rangle } \left(
  J_{\boldsymbol{x}\boldsymbol{y}} \vec S_{\boldsymbol{x}} \cdot \vec
  S_{\boldsymbol{y}} + \vec S_{\boldsymbol{x}}\cdot
  D_{\boldsymbol{x}\boldsymbol{y}} \vec S_{\boldsymbol{y}} \right),
  \label{HAMILTONIAN}
\end{equation}
($\langle \boldsymbol{x},\boldsymbol{y}\rangle$: lattice
nearest-neighbors). The random exchange-couplings,
$J_{\boldsymbol{x}\boldsymbol{y}}$, are Gaussian distributed with
$\overline{J_{\boldsymbol{x},\boldsymbol{y}}}=0$, and
$\overline{J^2_{\boldsymbol{x},\boldsymbol{y}}}=1$. The random
$D_{\boldsymbol{x}\boldsymbol{y}}$ are $3\times 3$ symmetric matrices
(i.e. $\vec S_{\boldsymbol{x}}\cdot D_{\boldsymbol{x}\boldsymbol{y}}
\vec S_{\boldsymbol{y}}= D_{\boldsymbol{x}\boldsymbol{y}} \vec
S_{\boldsymbol{x}}\cdot\vec S_{\boldsymbol{y}}$). Their matrix
elements are independent and uniformly distributed in $(-D,D)$. In
most of the work reported here \hbox{$D=0.05$} (which corresponds to
the best studied case~\cite{MATSUBARA}), but we will be presenting
results for \hbox{$D=0.1$} as well.

The ideal limit of a fully isotropic Heisenberg model is recovered
from Eq. (\ref{HAMILTONIAN}) by setting $D=0$. Once $D>0$, the
original $\mathrm{O(3)}$ symmetry, corresponding to a global spin
rotation (or reflection), is lost.  The only remaining symmetry for
$D>0$ is global spin inversion.

An instance of the couplings,
$\{J_{\boldsymbol{x},\boldsymbol{y}},D_{\boldsymbol{x},\boldsymbol{y}}^{\mu\nu}\}$
is named a {\em sample}. For any physical quantity, we first obtain
the thermal average, denoted as $\langle\ldots\rangle$. Only
afterwards we perform the sample average (denoted by an overline).

Defining the SG and CG susceptibilities requires real replicas.  We
consider pairs of spin configurations, $\vec S_{\boldsymbol{x}}^{a}$
and $\vec S_{\boldsymbol{x}}^{b}$, that evolve with independent
thermal noise, under the same couplings and at the same temperature.
The spin-overlap field is \mbox{$q_{\boldsymbol{x}}= \vec
  S_{\boldsymbol{x}}^a \cdot \vec S_{\boldsymbol{x}}^b$}, while its
Fourier transform at wave vector $\boldsymbol k$, is \mbox{$\hat
  q_{SG}(\boldsymbol{k})= \sum_{\boldsymbol{x}} q_{\boldsymbol{x}}~
  \mathrm{e}^{\mathrm{i} \boldsymbol{k} \cdot \boldsymbol{x}}/N$}.  On
the other hand, the local chirality is defined as:
\begin{equation}
  \zeta_{\boldsymbol{x}\mu}= \vec{S}_{\boldsymbol{x}+\boldsymbol{e}_{\mu}} \cdot ( \vec{S}_{\boldsymbol{x}} \times
  \vec{S}_{\boldsymbol{x}-\boldsymbol{e}_{\mu}})\,,\ \mu=1,2,3\,,
\label{LOCALQUIRAL}
\end{equation}
where $\boldsymbol{e}_{\mu}$ is the unit lattice vector along the $\mu$
axis. From (\ref{LOCALQUIRAL}), the chiral overlap-field is
$\kappa_{\boldsymbol{x},\mu}=\zeta_{\boldsymbol{x},\mu}^{a}~\zeta_{\boldsymbol{x},\mu}^{b}$,
where the superindices $a$ and $b$ correspond to the replicas. Its Fourier
transform is $ \hat q_{CG}^\mu(\boldsymbol{k})= \sum_{\boldsymbol{x}}
\kappa_{\boldsymbol{x},\mu}~ \mathrm{e}^{\mathrm{i} \boldsymbol{k} \cdot
  \boldsymbol{x}} / N$.  

The wave-vector dependent susceptibilities are:
\begin{eqnarray}
\chi_\mathrm{SG}(\boldsymbol{k})=N~\overline{\langle |\hat q_{SG}(\boldsymbol{k})|^2
  \rangle}\,,\ 
\chi_\mathrm{CG}^\mu(\boldsymbol{k})=N~\overline{\langle |\hat
  q_{CG}^{\mu}(\boldsymbol{k})|^2 \rangle}.\label{eq_SUSCEPT}
\end{eqnarray}
The correlation length, either SG or CG, is~\cite{COOPER,VICTORAMIT}
\begin{equation}
\xi= \frac{1}{2 \sin(k_\mathrm{min}/2)} \left( \frac{\chi(\boldsymbol{0})}{\chi(\boldsymbol{k}_\mathrm{min})} - 1 \right)^{1/2},
\label{eq_xi}
\end{equation}
where $\boldsymbol{k}_\mathrm{min}=(2\pi/L,0,0)$ or permutations.\cite{FN2}


Our simulation algorithm combines heat-bath with microcanonical
overrelaxation.\cite{OVERRELAX} Both moves generalize
straightforwardly to the anisotropic case.\cite{FN3} The mixed
algorithm is effective for the isotropic Heisenberg SG~\cite{HSGWE,
  LEEYOUNG2, HSG-PIXLEYYOUNG, VIET-KAWAMURA, HSG-LVSAP} and for other
frustrated models.\cite{OVERRELAXSPIN}  Besides, we extrapolate to
nearby temperatures using a bias-corrected~\cite{BIAS} data
reweighting method.\cite{SPECTRALDENSITY} Most of our simulations
were carried out with $D=0.05$, see
Table~\ref{SIMUDETAILS}. Nevertheless, we did as well some work for
$D=0.1$, see Table~\ref{SIMUDETAILS-NEW}.

\begin{table}[t]
\begin{center}
\begin{tabular}{|c|c|c|c|c|c|c|}
\hline 
$T~\diagdown~L$  & $6$    & $8$    & $12$   & $16$   & $24$   & $32$\\ 
\hline
\hline
 $0.187$         & $1000$ & $1080$ & $1000$ & $1020$ & $1000$ & $1000$ \\
 $0.194$         & $1000$ & $1080$ & $1000$ & $1020$ & $1000$ & $-$    \\
 $0.200$         & $1000$ & $1080$ & $1060$ & $1020$ & $1000$ & $-$    \\ 
 $0.210$         & $1000$ & $1080$ & $1000$ & $1020$ & $1200$ & $1000$ \\
 $0.220$         & $1000$ & $1080$ & $1000$ & $1020$ & $1080$ & $-$    \\    
 $0.230$         & $1000$ & $1080$ & $1000$ & $1020$ & $1080$ & $1000$ \\
 $0.240$         & $1000$ & $1080$ & $1000$ & $1020$ & $1080$ & $-$    \\
 $0.250$         & $1000$ & $1040$ & $1000$ & $1020$ & $1000$ & $-$    \\
\hline 
EMCS$\times10^{5}$ & $3$    & $3$    & $1.8$  & $3.6$  & $4.8$ & $15$\\
\hline
\end{tabular}
\end{center}
\caption{{\bf Details of simulations with $D=0.05$.} For each lattice size and
  temperature, we give the number of simulated samples. The last row
  indicates the number of Elementary Monte Carlo Steps (EMCS). The
  $L$-dependent EMCS consisted of 1 Heat-Bath full lattice sweep,
  followed by $5L/4$ {\em sequential} (microcanonical) overrelaxation
  sweeps. We took $6 \times 10^4$ measurements per sample, but for
  $L=32$ ($15 \times 10^4$ measurements).}
\label{SIMUDETAILS}
\end{table}

\begin{table}[t]
\begin{center}
\begin{tabular}{|c|c|c|c|c|}
\hline 
$T~\diagdown~L$  & $6$    & $8$    & $12$   & $16$ \\ 
\hline
\hline
 $0.230$         & $300$ & $100$ & $100$ & $320$  \\
 $0.240$         & $300$ & $100$ & $100$ & $260$  \\
 $0.250$         & $300$ & $100$ & $100$ & $360$  \\
 $0.260$         & $200$ & $500$ & $500$ & $820$  \\
 $0.270$         & $200$ & $400$ & $500$ & $560$  \\
 $0.280$         & $120$ & $600$ & $500$ & $500$  \\
\hline 
EMCS$\times10^{5}$ & $3$    & $3$    & $1.8$  & $3.6$ \\
\hline
\end{tabular}
\end{center}
\caption{As in Table~\ref{SIMUDETAILS}, for our simulations with $D=0.1$.}
\label{SIMUDETAILS-NEW}
\end{table}

\section{EQUILIBRATION}\label{SECT:EQUILIBRATION}

\begin{figure}[t]
\begin{center}
\includegraphics[width=\columnwidth]{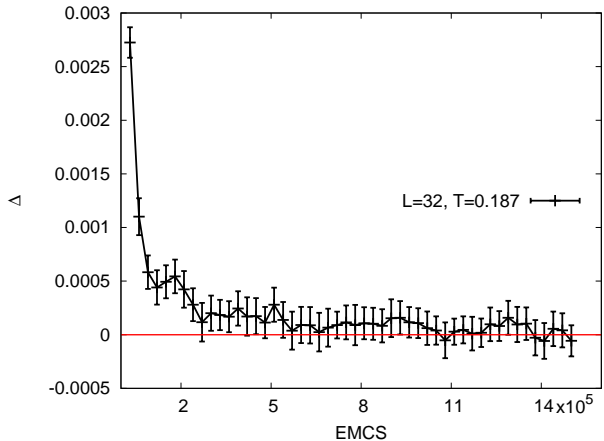}
\caption{(color online) Sample-averaged $\varDelta$ defined in the
  l.h.s. of Eq.~(\ref{equiltest}) vs. Monte Carlo time, as computed
  for $L=32$ at $T=0.187$ and $D=0.05$. The EMCS was defined in the
  caption to Table~\ref{SIMUDETAILS}.  Each point is an average over
  3000 consecutive measurements.  }
\label{fig_YSD_ZERO}
\end{center}
\end{figure}

We considered three thermalization tests. {\em First}, consider the
identity (valid for Gaussian-distributed
$J_{\boldsymbol{x},\boldsymbol{y}}$):
\begin{equation}
  \varDelta \equiv \overline{\frac{q_{s} - q_{l}}{T} +  \frac{2}{z}\, U}= 0,
  \label{equiltest}
\end{equation}
where $U=- \sum_{\langle \boldsymbol{x},\boldsymbol{y} \rangle}
J_{\boldsymbol{x},\boldsymbol{y}} \langle \vec S_{\boldsymbol{x}}
\cdot \vec S_{\boldsymbol{y}} \rangle /L^D$, the link-overlap is
$q_{l}=2\sum_{\langle \boldsymbol{x},\boldsymbol{y} \rangle} \langle
\vec S^a_{\boldsymbol{x}} \cdot \vec S^a_{\boldsymbol{y}} \rangle
\langle \vec S^b_{\boldsymbol{x}} \cdot \vec S^b_{\boldsymbol{y}}
\rangle /(z L^D)$, while $q_{s} = 2\sum_{\langle
  \boldsymbol{x},\boldsymbol{y}\rangle}\langle (\vec S_{\boldsymbol x}
\cdot \vec S_{\boldsymbol y})^2 \rangle/(z L^D)$ ($z=6$ is the lattice
coordination number). Now, both $U$ and $q_s$ equilibrate easily.
Yet, since $q_l$ involves {\em two} replicas, it slowly grows from
zero until its equilibrium value. Thus, a thermalization bias shows up
as $\varDelta>0$.\cite{LEEYOUNG2, HSG-LVSAP, KATZPALYOUNG}  The time
evolution of $\varDelta$, for $L=32$ at the lowest $T$, is in
Fig.~\ref{fig_YSD_ZERO}.  {\em Second,} we carried out the standard
logarithmic data binning: we compare averages over the second half of
the Monte Carlo history, with the second fourth, the second eight, and
so forth, finding stability for three bins. {\em Third,} we checked
for compatibility among reweighting extrapolations for contiguous
temperatures (our simulations at different $T$ are statistically
independent, see Fig.~\ref{fig_XICG_XISG}).

\section{RESULTS}\label{SECT:RESULTS}

\begin{figure}[t]
\begin{center}
\includegraphics[width=\columnwidth]{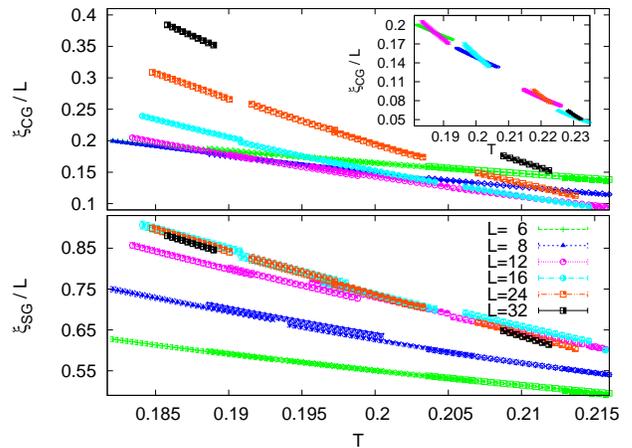} 
\caption{(color online) Correlation length in units of the lattice
  size vs. $T$, for all our system sizes at $D=0.05$. Results for both the CG
  ({\bf top}), and the SG sectors ({\bf bottom}). Data {\em patches}
  correspond each to an independent simulation (we used data
  reweighting~\cite{SPECTRALDENSITY}).  {\bf Inset:} CG intersections
  for pairs of sizes $(L, 2L)$. The range of $T$ and
  $\xi_\mathrm{CG}/L$ differ from main plot.}
\label{fig_XICG_XISG}
\end{center}
\end{figure}

Our FSS analysis compares the correlation length in units of the
lattice size for pairs of lattices
$(L,2L)$.\cite{BALLESTEROS,VICTORAMIT, QUOTIENTS} Dimensionless
quantities, such as $\xi/L$, are functions of
$L^{1/\nu}(T-T_\mathrm{c})$, $\nu$ being the thermal critical
exponent. Thus, the two curves intersect at $T_\mathrm{c}(L,2L)$, see
Fig.~\ref{fig_XICG_XISG}. $T_\mathrm{c}(L,2L)$ differs from
$T_\mathrm{c}$ due to scaling corrections (but tends to it for large
$L$~\cite{VICTORAMIT}). Our dimensionless quantities
$\xi_\mathrm{SG}/L$ and $\xi_\mathrm{CG}/L$, produce two
$L$-dependent critical temperatures $T_\mathrm{SG}(L,2L)$ and
$T_\mathrm{CG}(L,2L)$.  We compute the anomalous dimensions $\eta$
from the scaling of the susceptibilities $\chi$ [take
  $\boldsymbol{k}=\boldsymbol{0}$ in Eq.~(\ref{eq_SUSCEPT})]. For
large $L$, and $\eta<2$, $\chi$ diverges as
$\chi\propto|T-T_\mathrm{c}|^{-\nu(2-\eta)}$. For finite $L$, we
consider the susceptibility ratio for $\chi_\mathrm{CG}$ and
$\chi_\mathrm{SG}$ (the dots stand for scaling corrections):
\begin{equation}
\left.\frac{\chi(2L)}{\chi(L)}\right|_{T_\mathrm{c}(L,2L)}=2^{2-\eta}\,+\ldots\,.
\label{COCIENTES}
\end{equation}

\begin{table}[t]
\begin{center}
\begin{tabular}{|c|c|c|c|c|}
\hline $L$ & $T_\mathrm{SG}(L,2L)$ & $T_\mathrm{CG}(L,2L)$ & $2-\eta_{\mathrm{SG}}$ & $2-\eta_{\mathrm{CG}}$ \\ 
\hline
$6 $  & $0.251( 2)$ & $0.187(1)$ & $2.031(11)$ & $0.360(8)$\\
$8 $  & $0.235( 1)$ & $0.202(1)$ & $2.131( 9)$ & $0.223(8)$\\
$12$  & $0.207( 5)$ & $0.221(1)$ & $2.413(46)$ & $0.081(5)$\\
$16$  & $0.179(10)$ & $0.233(1)$ & $2.639(55)$ & $0.030(5)$\\
\hline
\end{tabular}
\end{center}
\caption{Size-dependent critical temperatures $T_\mathrm{SG}(L,2L)$
  and $T_\mathrm{CG}(L,2L)$, and anomalous dimensions
  $2-\eta_{\mathrm{SG}}$ and $2-\eta_{\mathrm{CG}}$,
  Eq.~(\ref{COCIENTES}), for the simulations with $D=0.05$.  Errors
  were obtained with jackknife.}
\label{TABLE_INTER-ETA}
\end{table}

We discuss first the CG sector. The inset of Fig.~\ref{fig_XICG_XISG} shows an
unusual feature: $\xi_\mathrm{CG}/L$ at the crossing point
$T_\mathrm{CG}(L,2L)$ approaches zero for large $L$. This is to be expected
only if $\eta\geq 2$:\cite{VICTORAMIT} if the susceptibility does not diverge
at $T_\mathrm{c}$, the correlation length in Eq.~(\ref{eq_xi}) scales as
$\xi/L \sim L^{-(\eta-2)/2}$. Nevertheless, we still find crossings when
comparing lattices sizes $L$ and $2L$, see Fig.~\ref{fig_XICG_XISG} and also
Ref.~\onlinecite{FUNNY-SCALING}. Crossings are due to the fact that, in the
large-$L$ limit, the correlation length in Eq.~(\ref{eq_xi}) {\em is}
divergent in the low-temperature phase. For $T<T_\mathrm{c}$, $\xi/L$ grows as
$L^{\theta/2}$ (i.e. the correlation function at large distances $r$ goes to a
constant with corrections of order $1/r^\theta$, see
e.g. Ref.~\onlinecite{JANUS-DECAY}).  Yet, the susceptibility ratio in
Eq.~(\ref{COCIENTES}) is constant for large $L$, even if $\eta>2$. So,
$\eta_\mathrm{CG}$ in Table~\ref{TABLE_INTER-ETA} approaches $2$ as $L$ grows.

Besides, it is note worthy that, in spite of the smallness of $D$,
$T_\mathrm{CG}(L,2L)$ for $D=0.05$ is about twice its value for the isotropic
model, $T_\mathrm{CG}(D=0)\approx 0.13$.\cite{HSG-LVSAP} In fact,
extrapolating the data in Table~\ref{TABLE_INTER-ETA} as
$T_\mathrm{CG}(L,2L)=T_\mathrm{CG}+A/L$ yields $T_\mathrm{CG}\approx 0.26$.

\begin{figure}[t]
\begin{center}
  \includegraphics[width=\columnwidth]{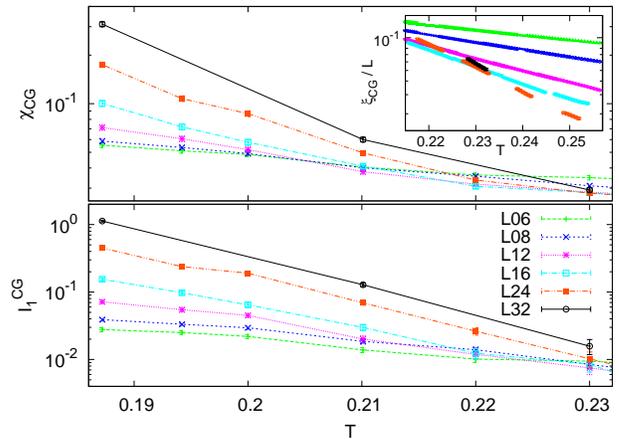}
  \caption{Chiral glass susceptibility $\chi_{CG}$ ({\bf top}) and
    $I_1$ integral defined in Eq.~\ref{Ik-def} ({\bf bottom})
    vs. temperature, for all our system sizes with $D=0.05$. Lines are
    guides to eyes. {\bf Inset}: zoom of $\xi_\mathrm{CG}/L$ vs. T}
  \label{fig_CHICG_I1}
\end{center}
\end{figure}

To further investigate the lacking divergence of $\chi_\mathrm{CG}$ at
$T_\mathrm{CG}$, we consider the integrals~\cite{JANUS-DYNAMICS}
\begin{equation}
I_k=\sum_{r=0}^{L/2} r^k C_\mathrm{P,P}(r)\,,\label{Ik-def}
\end{equation}
where $C_\mathrm{P,P}(r)$ is the plane-to-plane correlation
function.\cite{FN4} Note that $\chi_\mathrm{CG} \sim 2 I_0$, which
means that plane-to-plane correlation functions decays with $r$ slower
than the standard point-to-point correlations by a factor
$r^{D-1}$. The scaling behavior of the integrals~(\ref{Ik-def}) is:
$I_k\sim constant$ in the paramagnetic phase, $I_k\sim L^{k+2-\eta}$
at $T_\mathrm{CG}$ (if $k+2-\eta>0$, otherwise it is $I_k\sim
constant$), and $I_k\sim L^{D+k}$ in the CG phase. We show in
Fig.~\ref{fig_CHICG_I1}--top our data for $\chi_\mathrm{CG}$ (which is
basically $ 2I_0$) and, in Fig.~\ref{fig_CHICG_I1}--bottom,
$I_1$. Note that for $T<0.22$ the two integrals are diverging with
$L$. On the other hand, for $T=0.22,0.23$, $I_1$ grows with $L$, while
$\chi_\mathrm{CG}$ does not, as expected for $2<\eta_\mathrm{CG}<3$.

The behavior of the SG sector is more conventional. A remarkable
feature in Fig.~\ref{fig_XICG_XISG} and Table~\ref{TABLE_INTER-ETA} is
the strong scaling corrections in $T_\mathrm{SG}(L,2L)$. We do not
consider it safe to extrapolate $T_\mathrm{SG}$ to its large-$L$
limit, as we are far from the asymptotic regime. The SG anomalous
dimension takes a negative value as $L$ grows (also found in the Ising
SG, see e.g.~\cite{BALLESTEROS}).

An intriguing feature is that $T_\mathrm{SG}$ seems smaller than
$T_\mathrm{CG}$. Indeed, see Fig.~\ref{fig_XICG_XISG}, at $T\approx
0.187$, where $\xi_\mathrm{SG}/L$ becomes $L$-independent,
$\xi_\mathrm{CG}/L$ is growing fast with $L$. Yet, three caveats
prevent us from considering this conclusion as definitive: (i) our
lattice sizes are still in a strong cross-over regime, hence the final
picture could change as $L$ grows, (ii) the behavior is rather
marginal, meaning a larger number of samples would be needed to
accurately locate $T_\mathrm{SG}$ (this is hardly surprising, given
the large value of exponent $\nu$, and the small $\theta$ exponent,
for $d=3$ Ising SG's), and (iii) when considering a larger anisotropy,
see below, the effect seems smaller.

\begin{figure}[t]
\begin{center}
  \includegraphics[width=\columnwidth]{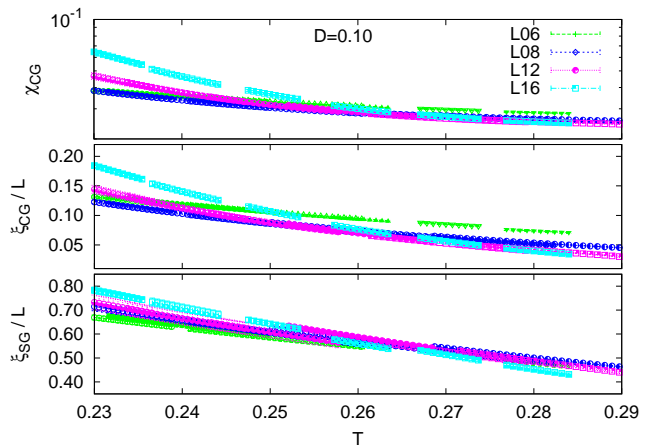}
  \caption{For all our lattice sizes at $D=0.1$, we show the
    chiral-glass susceptibility ({\bf top}), as well as the CG ({\bf
      center}) and SG ({\bf bottom}) correlation lengths in units of
    the system size, as a function of temperature.}
  \label{fig_CHICG_D010}
\end{center}
\end{figure}

Indeed, we have performed further simulations with $D=0.1$, up to
$L=16$. As shown in Fig.~\ref{fig_CHICG_D010}, the difference between
$T_\mathrm{SG}$ and $T_\mathrm{CG}$ is less clearly defined than for
$D=0.05$. On the other hand, the chiral-glass susceptibility is not
divergent at the critical point, in agreement with our results for
$D=0.05$. Consistently with that, the crossing points for
$\xi_\mathrm{CG}/L$ shift to a smaller height when $L$ grows.

\section{CONCLUSIONS}\label{SECT:CONCLUSIONS}
In summary, we have performed a finite size scaling study of the $3d$
Heisenberg spin glass in the presence of a weak random anisotropy, for
lattices of size up to $L=32$. Anisotropies cause that the CG susceptibility
no longer diverges at $T_\mathrm{CG}$, the chiralities ordering
temperature. Hence, the anisotropic system belongs to a Universality Class
different from the isotropic model (probably that of Ising SG's). Besides, we
found that the spin-glass ordering sets up only at
$T_\mathrm{SG}<T_\mathrm{CG}$. The most economic scenario is that actually
$T_\mathrm{SG}=T_\mathrm{CG}$ (the apparent difference would be due to
finite-size effects). In this scenario, chiralities would merely be a
composite operator (such as, say, the ninth power of the spin
overlap). However, the would-be intermediate temperature region where only
chiralities order should be experimentally detectable through the anomalous
Hall effect. Numerical studies covering a wider range of values for the
anisotropic coupling could also help to elucidate the situation.

\section*{ACKNOWLEDGMENTS}

We thank A. Tarancon, G. Parisi and P. Young for discussions. Simulations were
performed at BIFI ({\em Terminus}, $6.6\times10^5$ hours of CPU time) and Red
Espa\~nola de Supercomputaci\'on ({\em Caesaraugusta}, $2.79\times10^5$
hours), whose staff we thank for the assistance provided. We were partly
supported by MICINN (Spain) through research contract No. FIS2009-12648-C03,
and (S.P.G.)  through the FECYT Foundation.



\begin{thebibliography}{99}
\bibitem{FN2} We will not make any distinction between the CG longitudinal or
  transversal correlation lengths, see e.g. Ref.~\onlinecite{LEEYOUNG}, since
  the two turn out to be compatible within errors.
\bibitem{FN3} Eq.~(\ref{HAMILTONIAN}) can be split as $H=\vec
  h_{\boldsymbol{x}}\cdot \vec S_{\boldsymbol{x}} + H'$; $H'$ is independent
  of $\vec S_{\boldsymbol{x}}\,,$ while $\vec h_{\boldsymbol{x}}=\sum_{\Vert
    \boldsymbol{x}- \boldsymbol{y}\Vert=1} [J_{\boldsymbol{x},\boldsymbol{y}}
    \vec S_{\boldsymbol{y}}+ D_{\boldsymbol{x},\boldsymbol{y}} \vec
    S_{\boldsymbol{y}}]\,.$
\bibitem{FN4} For planes parallel to (say) the $\mu\!=\!3$ lattice axis, add
  the correlation function for all pairs of sites
  $\boldsymbol{x}=(n_1,n_2,n_3)$, $\boldsymbol{y}=(m_1,m_2,m_3)$ such that
  $m_3-n_3= r + k L$, with $k$ integer, then divide by $L^D$. We average over
  the three lattice directions, as well as over the longitudinal (weight 1/3)
  and the transversal (weight 2/3) correlation functions.
\end{thebibliography}
\end{document}